\documentstyle{rspublic}
\input{epsf.sty}

\newcommand{\gs}
           {\mathrel{\hbox{\rlap{\hbox{\lower4pt\hbox{$\sim$}}}\hbox{$>$}}}}
\newcommand{\ls}
           {\mathrel{\hbox{\rlap{\hbox{\lower4pt\hbox{$\sim$}}}\hbox{$<$}}}}

\newcommand{\minpoint}{\mbox{$'\mskip-4.7mu.\mskip0.8mu$}}

\newcommand{\sun}{\mbox{$\odot$}}
\def\ltsima{$\; \buildrel < \over \sim \;$}
\def\simlt{\lower.5ex\hbox{\ltsima}}
\def\gtsima{$\; \buildrel > \over \sim \;$}
\def\simgt{\lower.5ex\hbox{\gtsima}}

\def\arcm{$'~$}

\begin{document}

\title{Galaxy Clustering at z ${\bf \sim}$ 3}
\shorttitle{Structure at $z \sim 3$}
\author{C.~Steidel$^{1}$, K.~Adelberger$^{1}$, M.~Giavalisco$^{2}$, M.~Dickinson$^{3}$, M.~Pettini$^{4}$,
and M.~Kellogg$^{1}$} 
\shortauthor{C. Steidel \et}
\affiliation{(1) Palomar Observatory, California Institute of Technology,
(2) Observatories of the Carnegie Institution of Washington,
(3) Johns Hopkins University and Space Telescope Science Institute,
(4) Royal Greenwich Observatory}
\maketitle

\abstract 
Galaxies at very high redshift ($z \sim 3$ or greater) are now accessible to
wholesale observation, making possible for the first time a robust
statistical assessment of their spatial distribution at lookback times
approaching $\sim$90\% of the age of the Universe. This paper summarizes
recent progress in understanding the nature of these early galaxies,
concentrating in particular on the clustering properties. Direct comparison of the data
to predictions and physical insights provided by galaxy and structure formation models 
is particularly straightforward at these early epochs, and results
in critical tests of the ``biased'', hierarchical galaxy formation paradigm.  
\endabstract

\section{An efficient strategy for surveying the distant universe }

The last several years have witnessed an explosion in the quantity of
information available on the high--redshift universe, made possible
largely by new observational facilities such as the refurbished
Hubble Space Telescope and, particularly, the W. M. Keck 10m telescopes.
The result is that extremely distant galaxies have gone from elusive ``curiosities''
to common objects for which well--defined samples can be collected. For the
first time, real statistics are becoming available, allowing for empirical
insight into early galaxy and structure formation. As inherently interesting
as very high redshift galaxies are in there own right, since one is necessarily
observing galaxies close to the epoch of their formation, it is the ability
to {\it quantitatively} test the predictions of paradigms for galaxy and structure
formation with real data that will lead to significant progress in our overall understanding. 

In this paper, we discuss and summarize recent progress resulting from a survey
of very high redshift galaxies in which the selection of targets is somewhat more
complicated than the traditional method of limiting a sample by flux in a particular
passband; instead, we employ a selection whose primary purpose is to isolate
a reasonably well--defined sample of galaxies in a relatively small interval
of redshift.  
The motivation for employing a photometric culling process to separate likely high
redshift objects from the dominant foreground is that increasingly faint spectroscopic surveys selected
by apparent magnitude do not necessarily select distant objects with
very high efficiency (Cowie \et 1996); moreover, the well--known practical problems imposed
by the night sky background and the opacity of the atmosphere make it very
difficult to identify galaxies having redshifts larger than $z \simgt 1.3$, beyond which
there is a dearth of spectroscopic features that fall in the ``clean''
region of the optical window.  It has been recognized by many that 
it again becomes more straightforward to make positive spectroscopic identifications
at redshifts larger than $z \sim 2.5$, where the Lyman $\alpha$ transition and
a host of other relatively strong far--UV resonance lines enter the
ground-based window. The key to targeting exclusively 
the very high redshift galaxy population is to select on a
spectroscopic feature so dramatic that it is unmistakable
even in the very crude spectrophotometry afforded by broad-band imaging.
The natural choice for such a feature is the Lyman limit of hydrogen
at 912\AA\ (rest--frame), which enters far enough into the optical window
to be discerned based on ground-based photometry at $z \simgt 2.6$. This
spectral feature is expected to have contributions from the intrinsic
spectra of O and B stars, the Lyman continuum opacity of the galaxy in which
the stars are forming, and the statistical opacity of the neutral hydrogen in the
intergalactic medium; the net result is that the far--UV spectra of star--forming
objects should exhibit a precipitous drop--off to essentially zero intensity
near the rest--frame Lyman limit. 
For our own galaxy survey, we adopted a 3--band photometric system specifically tailored
to detecting this Lyman break in the vicinity of $z \sim 3$ (Steidel \&
Hamilton 1992, 1993). An illustration of how the 3 passbands would sample the far--UV
continuum of a galaxy near $z \sim 3$ is given in Figure 1.

\begin{figure}
\centering
\vspace*{5pt}
\parbox{\textwidth}{\epsfxsize=0.7\textwidth \epsfbox{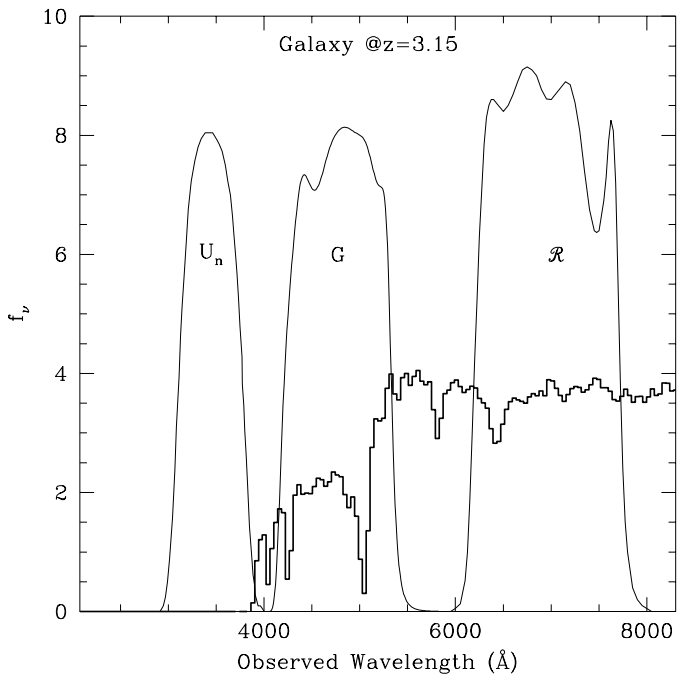}}
\caption{An illustration of how the adopted filter system is ``fine--tuned''
for 
observing the Lyman continuum break at $z \sim 3$. The model galaxy spectrum
includes the spectral energy distribution of the stars, but also includes
a reasonable component neutral hydrogen in the galaxy itself, and the statistical
effects of intervening neutral hydrogen (the dip in the spectrum just shortward
of Lyman $\alpha$ at 5050 \AA\ is due primarily to the line blanketing of the intervening
Lyman $\alpha$ forest. 
}
\end{figure}

\begin{figure}
\centering
\vspace*{5pt}
\parbox{\textwidth}{\epsfxsize=0.7\textwidth \epsfbox{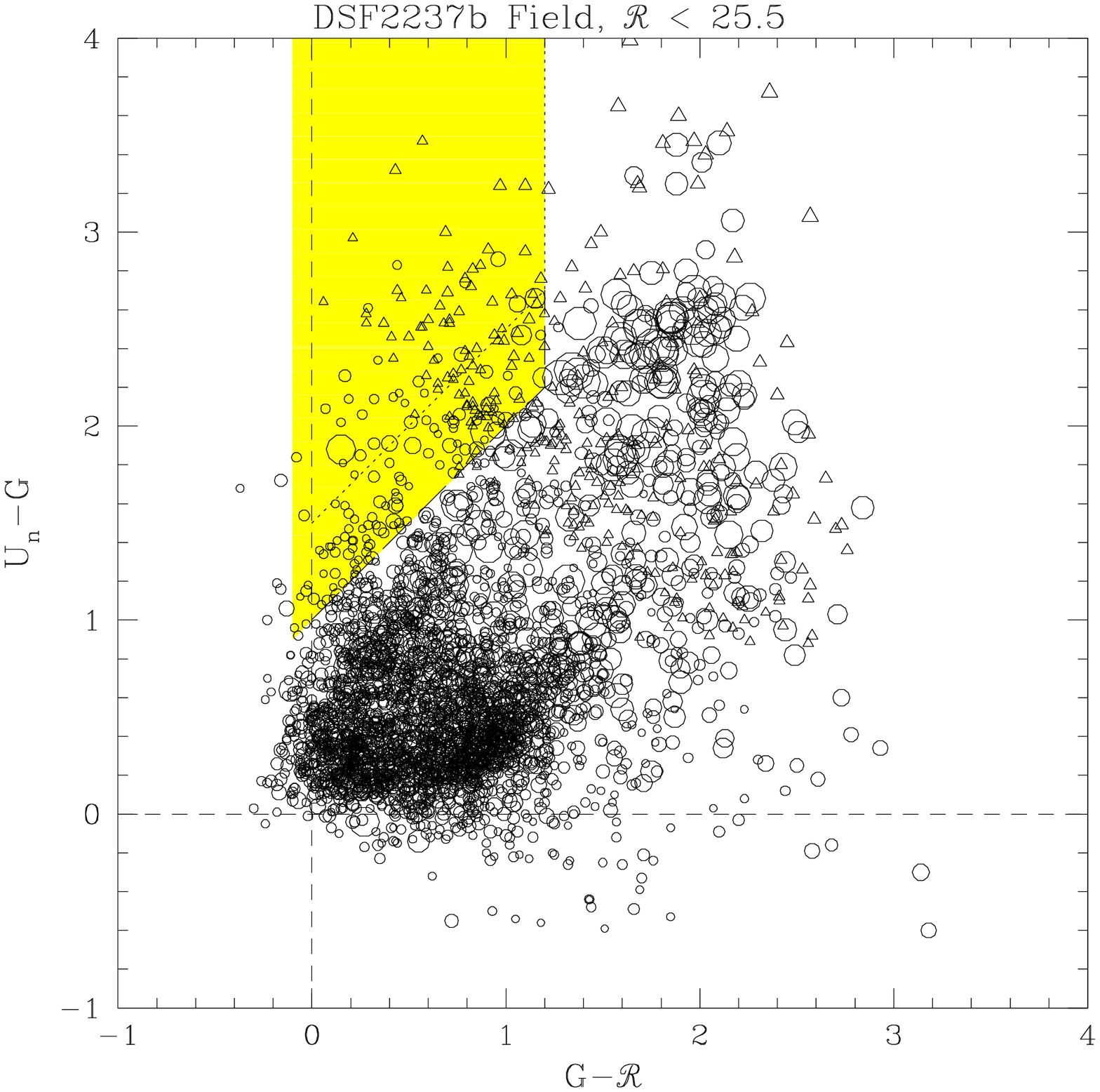}}
\caption{
A two--color diagram typical of those used to identify Lyman break
galaxy candidates for spectroscopic follow--up. The region of the color--color
plane populated by Lyman break galaxies in the redshift range $2.6 \simlt z \simlt 3.4$
is shaded. This example includes 3300 objects in a field of size 9\minpoint1 by 9\minpoint1;
a total of 140 of the objects to ${\cal R}=25.5$ satisfies the adopted color selection
criteria, or about 4\% of the total.}
\end{figure}

It is possible to make simple predictions of the spectral energy distributions
of distant galaxies, (e.g., Steidel, Pettini, \& Hamilton 1995)
based on modeling the far--UV spectra of star forming galaxies, and including 
the effects of both Lyman continuum opacity of the galaxy interstellar medium
and the known statistical effects of the intergalactic medium [see Madau
1995 for an in-depth discussion of the latter effect]. Based on such
predictions, one can isolate a region of "color-color space" in a diagram
such as that shown in Figure 2, in which {\it only} galaxies at $z > 2.6$
should be found. One would predict that a sample selected
from that region would have a redshift distribution that is limited on the low--redshift
side by the necessity of observing a significant
"break" across the Lyman limit in the fixed $U_n$ and $G$ passbands, and on
the high redshift side 
by the $G-{\cal R}$ color, which becomes increasingly
``reddened'' by the blanketing from the Lyman alpha forest. Both these effects
are rather easily modeled, and even before any confirming spectroscopy,
one might predict that the redshift range for objects in
the shaded region of Fig. 2 would be $2.7 \simlt z \simlt 3.5$.
In a sense this use of colors is akin to the increasingly
popular "photometric redshift" method, but our real intention is not
to measure redshifts with photometry, but to obtain something close to a 
a volume--limited (really, redshift--bounded) sample of galaxies
where the culling process would be highly efficient. Quite honestly,
even in our most optimistic times during several years of collecting
photometric data (see, e.g., Steidel, Pettini, \& Hamilton 1995) we
would not have imagined how cleanly this this method could be implemented 
with ground-based photometry of very faint galaxies.

It was our
first opportunity to use the Low Resolution Imaging Spectrograph 
(Oke \et 1995)
on the (then only) Keck telescope in September of 1995 that 
allowed us to convince ourselves and others that the method would
really work (Steidel \et 1996). It quickly became clear that it
would be feasible to construct large samples of $z \sim 3$ galaxies
with some concentrated effort; we thus began a project to obtain
images in our $U_nG{\cal R}$ photometric system of relatively large regions of sky, 
from which Lyman break candidates could be selected and followed up spectroscopically
on the W.M. Keck telescopes. The rationale for 
undertaking such a survey was that a large statistically homogeneous
sample was bound to be useful for a general understanding of the 
nature of the high redshift star forming galaxy population, and
it would almost certainly provide unprecedented information on the
clustering properties of very early galaxies, which one might
expect to provide a very sensitive cosmological test.

\begin{figure}
\centering
\vspace*{5pt}
\parbox{\textwidth}{\epsfxsize=0.8\textwidth \epsfbox{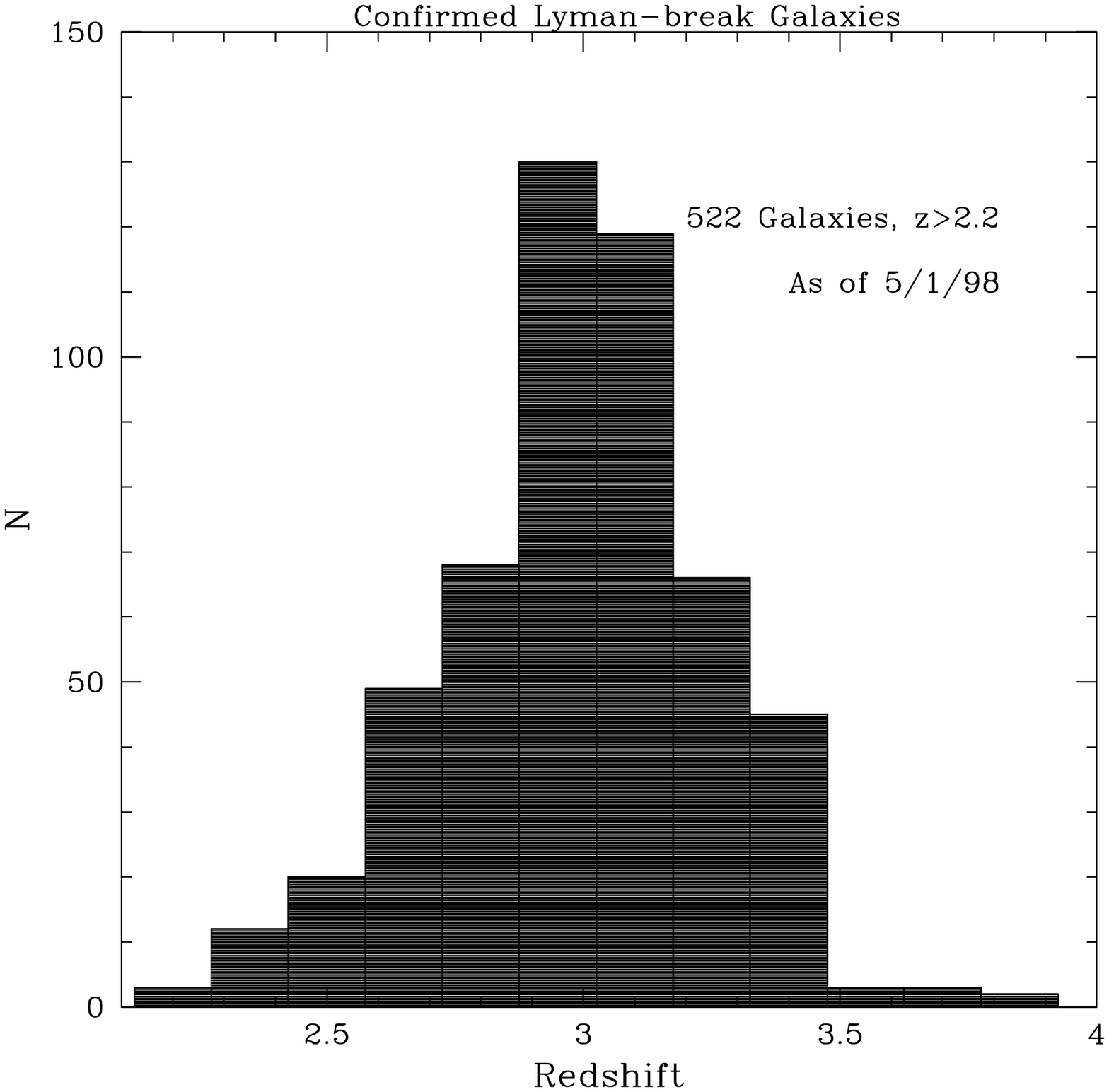}}
\vspace*{5pt}
\caption{The histogram of spectroscopically confirmed Lyman break galaxies
as of 1998 May. Note the well--defined sensitivity of the survey in
redshift space, fairly well-characterized by a Gaussian centered at $z=3.0$
with a dispersion $\sigma_z=0.25$.}
\end{figure}

\section{The Lyman Break Galaxy Survey}

The present goal of the LBG Survey is to cover 5-6 fields, each of
size 150--250 square arc minutes, for a total sky coverage of
about 0.3 square degrees. A typical survey field is 9\arcm\ by 18\arcm\,
so that the transverse co-moving scale is $\sim 12h^{-1} \times 24h^{-1}$ Mpc
for $\Omega_m=0.2$ open and $\Omega=0.3$ flat, and $\sim 8h^{-1} \times
16h^{-1}$ Mpc for $\Omega_m=1$; the effective survey depth is
$\sim 400h^{-1}$ Mpc for the low--density models and $\sim 250h^{-1}$ Mpc
for Einstein-de Sitter.   Within the full survey area, there will be 
approximately 1500 objects satisfying the color criteria illustrated
in Figure 2. The aim is to obtain confirming spectra for approximately
50\% or more of the photometric sample in the primary survey fields. The redshift
histogram of spectroscopically--confirmed objects at the time of this
writing (May 1998) is shown in Figure 3. Of these, 437 redshifts have
been obtained in what we now consider to be our primary survey fields.
To a large extent, the ``bottle--neck'' in the progress of the survey
is in obtaining the deep CCD images necessary for accurate photometric
selection; these images require approximately 2 clear nights on a 4-meter
class telescope per pointing, and most of our photometry has been obtained
at the prime focus of the Palomar 200--inch telescope, which provides a field
of only $\sim 9$\arcm\ square. A clear night with LRIS on the Keck II telescope
will typically yield 50-60 confirmed $z \sim 3$ galaxies, so that the entire
survey could in principle be completed after a total of 15-20 nights (we are
approximately 60\% finished at this time).

Figure 3 shows that the peak of the sensitivity of the survey lies
at $z = 3.02$, with about 90\% of the objects lying in the interval
[2.7,3.4].  The survey is obviously incomplete on either side
of the median redshift; to calculate the effective volume covered
by the survey we assume that it is 100\% complete at $z=3$ and that
the true LBG density does not change significantly over the range
of interest. To ${\cal R}=25.0$, the observed surface density of
Lyman break galaxies satisfying the color criteria illustrated in Figure 2 
is 1.0 per square arc minute, corresponding to co--moving space densities
of $6.4\times10^{-3}h^{3}$ Mpc$^{-1}$ ($\Omega_m=1)$ or $1.7\times10^{-3}h^3$ Mpc$^{-3}$
for either $\Omega_m=0.2$ open or $\Omega_m=0.3$ flat. For an Einstein-de Sitter
Universe, the space density integrated to ${\cal R}=25.0$ is roughly equivalent to the
present--day space density of galaxies with $L > L^{\ast}$; the density is 4 times smaller
than this for a universe with $\Omega_m \sim 0.2-0.3$.  Thus, the sample of Lyman
break galaxies represents relatively common objects, albeit objects at the bright
end of the far-UV luminosity distribution, and in the absence of severe censoring
by dust, these are the objects harboring the most vigorous star formation at
$z \sim 3$. 

Discussions of the far-UV luminosity function, the extinction corrections that
are likely to apply to the LBG population (and therefore the corrected star
formation rates), and the spectroscopic and morphological properties of the sample have been, or will
soon be, presented elsewhere (e.g., Pettini \et 1997, 1998; Dickinson 1998; Giavalisco 1998;
Steidel \et 1998b).

\begin{figure}
\centering
\vspace*{5pt}
\parbox{\textwidth}{\epsfxsize=0.7\textwidth \epsfbox{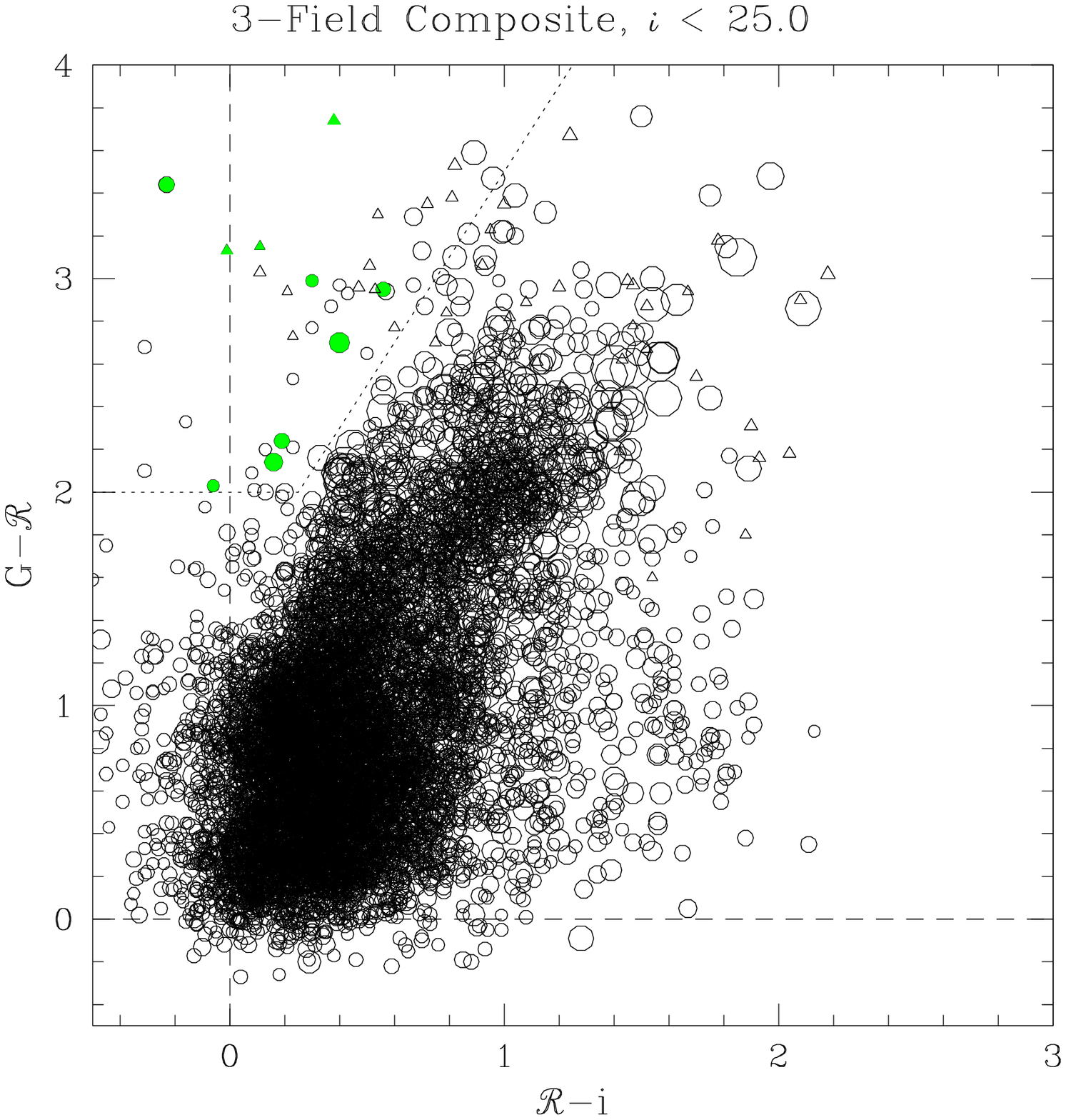}}
\parbox{\textwidth}{\epsfxsize=0.6\textwidth \epsfbox{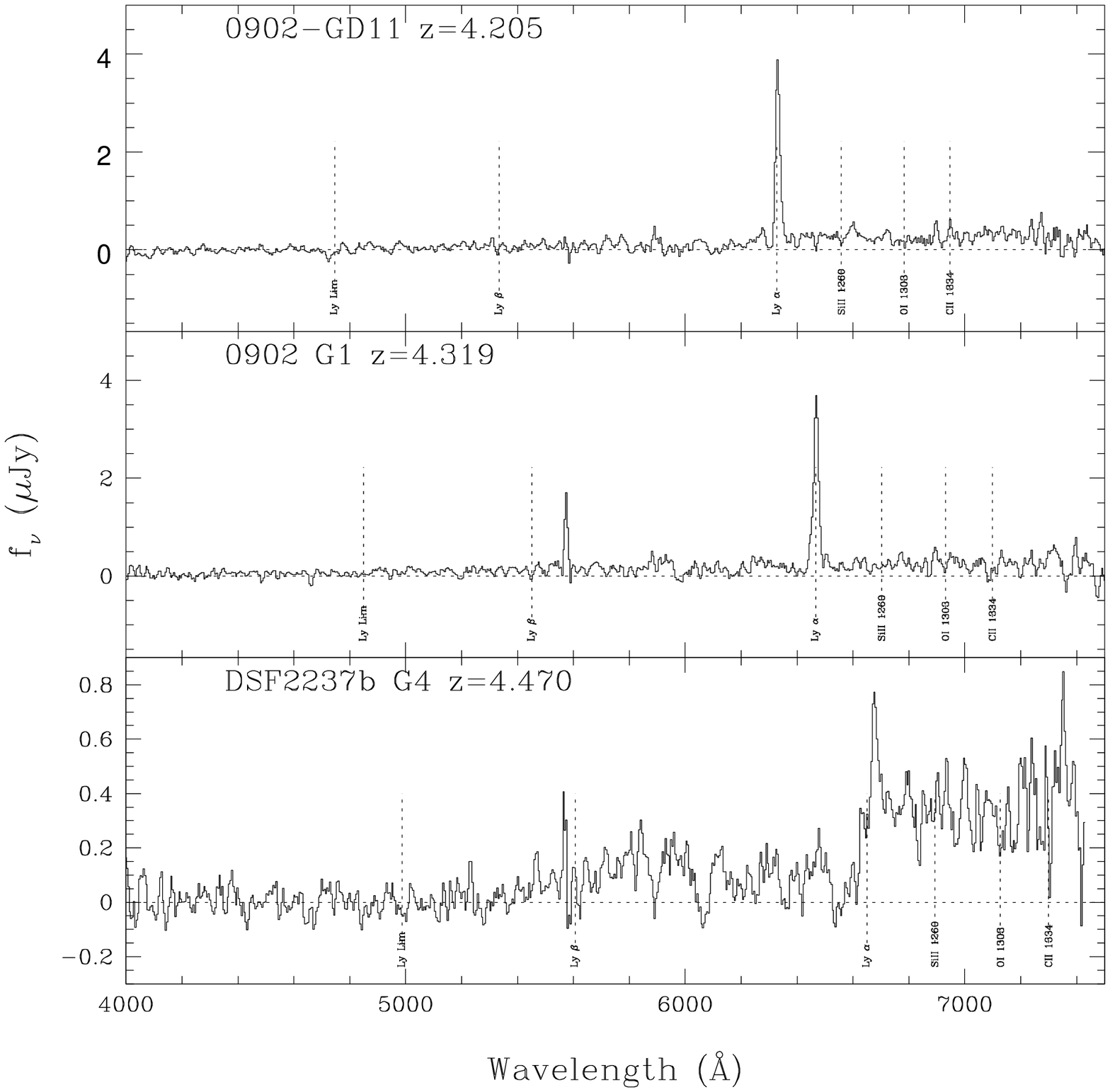}}
\vspace*{5pt}
\caption{The upper panel shows an example of a color--color diagram that
can be used to select galaxies in the range $3.9 \simlt z \simlt 4.5$
in a manner analogous to the $z \sim 3$ method. The filled symbols represent
objects which have been confirmed spectroscopically in the expected redshift
range.  The lower panel shows example spectra;
the first two clearly have very strong Lyman $\alpha$ emission, whereas
the third is much weaker (and unfortunately much more typical). Note the 
strong continuum break shortward of Lyman $\alpha$ emission due to the onset of the Lyman $\alpha$
forest. }
\end{figure}

An obvious extension of the current Lyman break selection technique is to
move the method to higher redshifts using a different filter system.
It has been straightforward to obtain data in one additional passband,
$i$[8100/1200], in our survey fields, so that one can search for objects exhibiting ``breaks''
in the $G$ band rather than the $U_n$ band. Models similar to those used for
defining the initial color cuts for $z \sim 3$ galaxies can be used to predict that,
for the color criteria defined in Figure 4a, the range of redshifts should
be $3.9 \simlt z \simlt 4.5$, for an expected median redshift of $z \sim 4.2$.
Our spectroscopic sample in this redshift range is still relatively small (example
spectra of $z > 4$ Lyman break galaxies are shown in Figure 4b), but not surprisingly 
the ``predictions'' are largely borne out. What is clear from our experimentation
with the $z \sim 4.2$ samples is that a large survey aimed at establishing the large-scale
distribution at this higher redshift interval would be {\it much} more difficult
that at $z \sim 3$. The reason for this is almost completely practical--- at $z \sim 3$,
all of the spectroscopic features useful for redshift identification fall
comfortably in the 4500--6500 \AA\ range, where the sky background is very dark,
the instrumental throughput is at a maximum, and there is
no fringing of the CCD which severely compromises one's ability to do precision
sky subtraction at longer wavelengths. At $z \sim 4.2$, the same features have
moved into the $6300-8000$ \AA\ range, where the sky is much brighter and
sky subtraction much more subject to systematic difficulties produced by fringing
and the ``forest'' of OH emission lines in the sky. As a result, the efficiency
with which one can go from photometric candidates to spectroscopic confirmations is
down by a factor of $\sim 5$, and it becomes especially difficult to confirm
objects without strong Lyman $\alpha$ emission lines.  For this reason,
we do not intend any major galaxy survey at $z \sim 4.2$, but our aim instead
is to establish the redshift selection function in order to make a statistically
significant differential comparison of the space density of star-forming galaxies
at $z \sim 4.2$ relative to those at $z \sim 3$, as we regard it as very important
to check the result implied in the Hubble Deep Field (Madau \et 1996) that the
space density of Lyman break galaxies is significantly lower at $z \sim 4$
than at $z \sim 3$.

In parallel with the large spectroscopic survey, we are also pursuing programs
involving near--IR imaging of sub-samples using Keck/NIRC, observations in the
sub--mm continuum of the most apparently reddened examples of $z \sim 3$ LBGs
using SCUBA on the JCMT, near--IR spectroscopy in order to obtain line widths
and fluxes of rest--frame optical nebular lines using UKIRT+CGS4 (Pettini \et 1998), and higher-dispersion
optical spectroscopy of selected bright examples using LRIS on Keck. Since most of
these investigations are related more to the astrophysics of the individual galaxies,
rather than their large-scale distribution, we will not discuss the results further in
the present summary. 

\section{Large Scale Structure at $z \sim 3$}

It was quite obvious (even at the telescope) during our first observing runs
spent collecting significant numbers of Lyman break galaxy spectra over
relatively large fields that the redshifts were far from randomly distributed
throughout the survey volume. Strong redshift--space clustering is certainly
not a new phenomenon for redshift surveys having ``pencil--beam'' geometries
(e.g., Broadhurst \et 1990, Cohen \et 1996); nevertheless, it was somewhat
surprising to encounter significant ``spikes'' in the redshift distribution
at $z\sim 3$, where naively one might expect clustering to be significantly
weaker than at $z < 1$ under any structure formation scenario that involves
gravitational instability. 

The first field for which a significant number of redshifts was obtained,
SSA22 (see the top left panel of Figure 5), yielded a structure on a
scale of $\sim 10$ Mpc that would be extremely rare 
for any cosmology (even for $\Omega_m=0.2$)
if galaxy number density fluctuations were an unbiased tracer of matter
fluctuations and if one adopted ``cluster normalization'' for the
value of $\sigma_8$ (e.g., Eke, Cole, and Frenk 1996). To have 
a significant probability of being found, a peak with the observed
over-density on the observed scale 
requires significant bias of the galaxy fluctuations as compared to underlying mass
fluctuations (Steidel \et 1998a). With a bias parameter on $\sim 10$ Mpc scales defined in the
usual way, $b\equiv \delta_{\rm gal} / \delta_{\rm mass}$, and assuming that such a peak
would be found in every survey field, a ``high peak'' analysis would require that
$b \simgt 6$ for an Einstein-de Sitter universe; the corresponding numbers would be
$b \simgt 2$ for $\Omega_m=0.2$ (open) and $b \simgt 4$ for $\Omega_m=0.3$ (flat). 
Our first reaction was that the very high galaxy bias required in the universe with
$\Omega_m=1$ was {\it too} high, and that this favored a low--density universe.
However, it turned out that such large values of the bias emerge naturally
for rare dark matter halos that are just collapsing at the epoch corresponding to
$z \sim 3$, within the context of CDM--like models for both N-body simulations
(Jing \& Suto 1998; Bagla 1998; Wechsler \et 1998; Governato \et 1998) and for 
analytic variations of Press-Schechter
theory (Press \& Schechter 1974; Mo \& Fukugita 1996; Mo \& White 1996; Baugh \et 1998).
It was also interesting, as we had remarked, that if a similar large peak were found
in each survey field, they would have just about the right space density to match
that of present--day X-ray clusters, suggesting the possibility that the ``spike'' could
be a proto--cluster viewed prior to collapse and virialization (there was no evidence
for central concentration of the galaxies within the ``spike'' on the plane of the sky). 
This interpretation is indeed supported by the simulations (Wechsler \et 1998, Governato \et 1998).
In any case, despite the frustrating result that strong clustering was expected for
the most massive virialized halos at $z \sim 3$ in {\it any} hierarchical model, it was
clear that the general paradigm of ``biased'' galaxy formation (Kaiser 1984; Bardeen \et 1986; Cole \& Kaiser 1989)
was strongly supported. Nevertheless, the numbers were quite uncertain based on
a single high peak in a single survey field, and it was clearly essential to obtain
more data so that the galaxy fluctuations could be better--characterized.

\begin{figure}
\centering 
\vspace*{5pt}
\parbox{\textwidth}{\epsfxsize=\textwidth \epsfbox{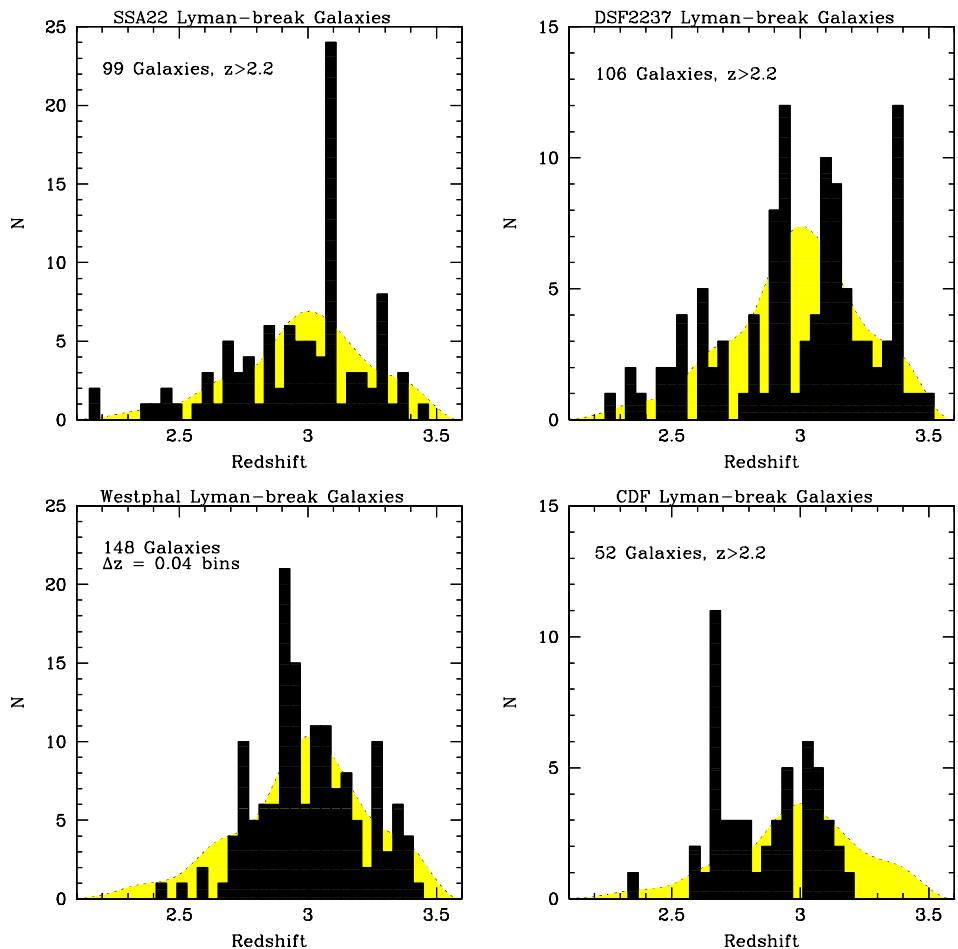}}
\vspace*{5pt}
\caption{Redshift histograms (to date) in 4 of our survey fields to date. The transverse
size of the surveyed fields varies: 8\minpoint7 by 8\minpoint7
(CDF),  9\arcm\ by 18\arcm\ (DSF2237, SSA22), 
15\minpoint0 by 15\minpoint0 (Westphal). In each case, the light histogram indicates
the empirical redshift selection function imposed by the photometric selection, normalized
to the same number of objects as observed. Note the prominent redshift ``spikes'' present
in each field.}
\end{figure}

Figure 5 contains redshift histograms from 4 of our survey fields, showing that
large fluctuations are indeed generic. To make this more quantitative, we have recently
analyzed the counts-in-cells fluctuations of LBGs within
six 9\arcm\ by 9\arcm\ fields in which the spectroscopy is reasonably complete (Adelberger \et 1998). This type
of analysis, which takes into account not just the highest peak, but general fluctuations
on a fixed co-comoving scale, should provide a much more robust estimate of the effective
bias of the LBGs.  The cells were cubes of side length defined by the transverse size of the field, or
$\sim 8h^{-1}$ Mpc for $\Omega_m=1$ and $\sim 12h^{-1}$ Mpc for $\Omega_m=0.2$ open
and $\Omega_m=0.3$ flat models. After correcting for shot noise, 
we found 
that $\sigma_{\rm gal} = 1.1 \pm 0.2$, implying that $b \equiv \sigma_{\rm gal}/\sigma_{\rm mass}$
is $6 \pm 1$, $1.9 \pm 0.4$, and $4.0\pm 0.7$ for $\Omega_m=1$, $\Omega_m=0.2$ open, and
$\Omega_m=0.3$ flat. These numbers are in very good agreement with our initial estimate from
a single high peak in the first field observed. If these inferred bias values are
used to estimate the more familiar galaxy--galaxy correlation length $r_0$, then for
a power law slope $\gamma=-1.8$ for the correlation function, the co-moving
correlation length would be $r_0=4h^{-1}$, $6h^{-1}$, and $5h^{-1}$ Mpc for 
$\Omega_m=1$, $\Omega_m=0.2$ open, and $\Omega_m=0.3$ flat, respectively. Note
that these values are roughly the same as the correlation length for galaxies
today, indicating the very strong bias that must be present relative to the
mass distribution in any reasonable gravitational instability scenario (cf. 
Baugh \et 1998, who predicted similar correlation lengths for Lyman break 
galaxies using their
semi-analytic galaxy formation model).
The published correlation lengths for intermediate redshift galaxy samples are significantly
smaller (cf. Le F\`evre \et 1996, Carlberg \et 1997), illustrating that the correlation strength 
of galaxy samples is almost certainly
strongly dependent on redshift and sample selection method in ways that are not
normally accounted for in simple models (see Giavalisco \et 1998a).

\begin{figure}
\centering
\vspace*{5pt}
\parbox{\textwidth}{\epsfxsize=0.8\textwidth \epsfbox{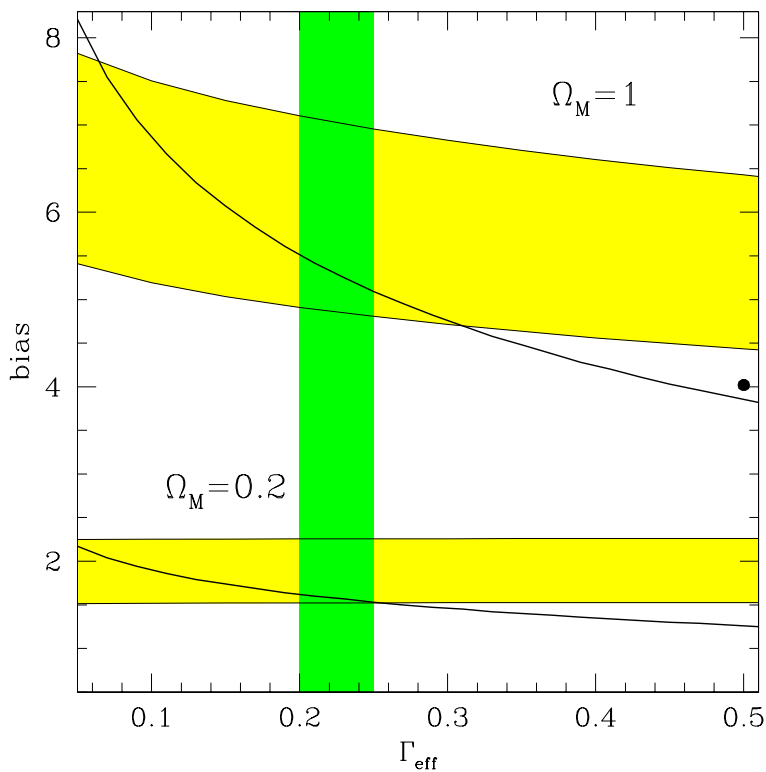}}
\vspace*{5pt}
\caption{A plot comparing the abundance and clustering properties of observed
galaxies and theoretical dark matter halos, from Adelberger \et 1998.  The 
horizontal shaded regions show the inferred bias (on $\sim 10$ Mpc scales) for observed
LBGs, versus a parameterization of the shape of the
mass fluctuation power spectrum. Analyses of present--day large scale
structure suggest a value of $\Gamma$ in the range $0.2 \simlt \Gamma \simlt 0.25$
(e.g., Peacock \& Dodds 1994),
shown with the vertical shaded region. 
(In this context, $\Gamma$ can be thought of as an indicator
of the relative power on galaxy [$\sim 1$ Mpc] and cluster [$\sim 10$ Mpc] scales).   
The solid curves show the predicted bias
of {\it halos} having the same abundance as the observed Lyman break galaxies. Note
the good agreement between the predictions and the observations if
$\Gamma \sim 0.2$,  and that standard CDM (the dark point) is discrepant by
about 2 $\sigma$.
}
\end{figure}

Once a reasonable estimate of the LBG bias is available, it is possible to make more
detailed comparisons to dark matter models. In particular, a successful model should be
able to produce simultaneously both the observed strong clustering of the LBGs, and
the right number density of halos exhibiting that strong clustering. 
The number density reflects the level of power on galaxy ($\sim 1$ Mpc) scales, while
the strong clustering we observe (e.g., in Figure 5) reflects power on $\sim 10$ Mpc
scales; a model will be able to match both observational constraints simultaneously
only if it has the right ratio of power on these two scales. This is illustrated
in Figure 6, where the ratio of power on these scales is parameterized in the
usual way with the power spectrum ``shape parameter'',  $\Gamma$. Higher values of
$\Gamma$ correspond to larger ratios of small to large scale power, and
(as explained in Adelberger \et 1998) to weaker clustering for objects of fixed 
abundance. $\Gamma \simlt 0.2$ is apparently required to reconcile the dark matter
model and the LBG observations; similar values are implied by observations of
galaxy clustering on scales $> 10$ Mpc in the local universe.   
Both the theoretical
and observational estimates of $b$ in Figure 6 are based on the same cluster normalization for
$\sigma_8$, so that changing the normalization will move the theoretical curve
and the empirical estimates of $b$ in much the same way (this explains why the shape of the
curves in Figure 6 are very similar for very different values of $\Omega_m$). 
The most important point to glean from Figure 6 is that 
one can match both the abundance {\it and} the clustering properties (parameterized
here by the value of $b$ on $\sim 10$ Mpc scales) of dark matter halos and the
observed galaxies using a simple model, provided that the shape of the power spectrum
is in the same range implied by local estimates of large
scale structure. 

An additional test of a generic hierarchical model would be that more abundant
objects must be less strongly clustered (i.e., less massive halos must exhibit
smaller values of $b$). Figure 7 shows the predictions of $b$ versus abundance
for a model having $\Gamma=0.2$. Again, there is no fitting involved here,
and it can be seen that in fact the much more abundant, much fainter LBGs
from the Hubble Deep Field sample are in fact much less strongly clustered,
entirely consistent with the predictions of the simple model (see Giavalisco
\et 1998b for a complete description of the models and of the HDF sample). Also of note is that
models with low $\Omega_m$ and Einstein-de Sitter models are equally capable
of matching the observations, given a spectral shape fixed
at $\Gamma =  0.2$; in both models, the relatively rare peaks in the density
field are expected to be strongly clustered (although the bias relative to
the overall mass distribution is very different) 
and to have roughly the same
dependence on halo number density. A very large difference, however, exists
in the predicted mass scales for the most strongly biased dark matter halos.
In the $\Omega=0.2$ model, the characteristic mass of halos having the
abundance (and clustering properties) of the spectroscopic LBG sample is $\sim 3\times 10^{12}$ M$_{\sun}$,
whereas the predicted mass of the same objects in the $\Omega_m=1$ model
is only $1.3 \times 10^{11}$, a difference of more than a factor of 20!
While dynamical mass estimates of these high redshift galaxies are extremely
challenging (see Pettini \et 1998), the differences are so large that
it may be quite plausible to discriminate observationally between the two cosmologies. 

\begin{figure}
\centering
\vspace*{5pt}
\parbox{\textwidth}{\epsfxsize=0.8\textwidth \epsfbox{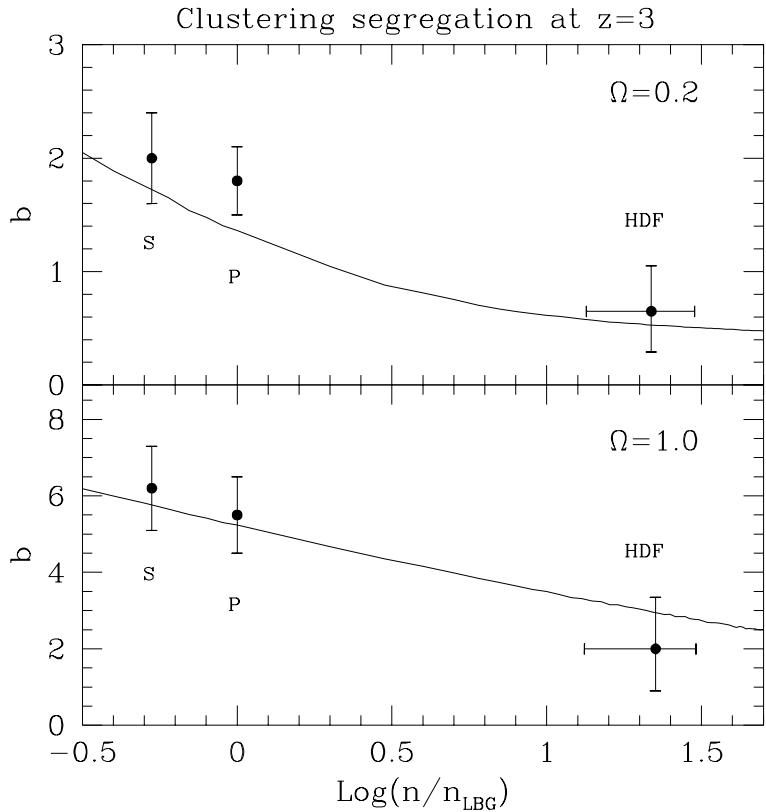}}
\vspace*{5pt}
\caption{The predicted bias as a function of halo abundance for a 
dark matter model with $\Gamma=0.2$, together with data from
3 different samples of LBGs. The point marked "S" is based on the
spectroscopic sample discussed above; the point marked "P" is from a 
larger and slightly fainter photometric sample for which the clustering properties were
estimated using the observed angular correlation function together
with the observed redshift selection function (Giavalisco \et 1998a). 
The point marked "HDF" comes from a $w(\theta)$ analysis of F300W
"dropouts" to $F606W=27$ in the Hubble deep field; the error bars
reflect the uncertainties due to the fact that the redshift selection
function for the HDF color selection criteria are not precisely known. 
Note that no "fitting" of the model curve has been imposed. 
}
\end{figure}

\section {What Does It All Mean?}

The data are obviously just reaching the level where quantitative analyses are
possible, and there is no doubt that the observational situation can
be improved dramatically on a short timescale. However, based on what must
be considered preliminary analysis of the current data, it is already
possible to list some broad conclusions that are unlikely to change
substantially.                               

First, the clustering properties of Lyman break galaxies, which are selected
on the basis of their rest--frame far-UV flux, indicate that they are associated
with relatively rare, massive dark matter halos. This is true independent of
the matter density; however, for a power spectrum shape that obeys local
constraints ($\Gamma \sim 0.2$), the mass scale associated with the most luminous
LBGs is strongly $\Omega_m$--dependent. For low--density models, the
halo mass scale is $\sim 10^{12}$ M$_{\sun}$, already similar to massive
galaxies at the present epoch. The strong clustering of massive halos is
expected for standard hierarchical models in which the fluctuations are
Gaussian, and thus the LBGs are apparently tracing regions of enhanced mass density
at early epochs. In the context of models of hierarchical growth of structure, 
this means by and large that the descendents of LBGs would be found as parts
of much larger virialized structures in the universe today (e.g., Steidel \et 1998;
Governato \et 1998; Wechsler \et 1998). The strongest peaks in the distribution of
LBGs at high redshift are likely to be the progenitors of rich clusters of galaxies,
which one is apparently seeing prior to collapse and virialization.  
Regardless of the details of one's interpretation, the ``paradigm'' that galaxies
form at the (biased) high peaks in the dark matter distribution is very strongly
supported by the data. 

The statistics are now good enough that an attempt to reconcile the abundance
and clustering properties of LBGs with models is justified. Quite remarkably
(in our opinion), there is amazingly good agreement between the predictions of
a simple dark matter model having the power spectrum shape constrained by
local large scale structure, and the observed galaxies. As discussed in Adelberger \et 1998, this
agreement depends on a very tight relationship between dark matter halo
mass and far--UV luminosity, as this is implicit in matching observed galaxies
to dark matter halo abundances. If it were the case that star formation were a highly
stochastic process, in which halos differing substantially in mass could produce
the same star formation rate, it would ``dilute'' the clustering properties
of a sample selected by UV luminosity so that it would not result in 
clustering as strong as observed. Further, it is difficult to reconcile the models
and the data unless there is essentially a one-to--one correspondence between {\it observable}
galaxies and dark matter halos (if we were observing only a small fraction of
the strongly--clustered massive halos, then this would present a problem for
any hierarchical model). This, incidentally, argues against a large population of
star forming galaxies completely obscured by dust, and also against models in
which the LBGs are undergoing brief bursts of star formation that ``light up'' only a small fraction
of the halos at a time.  The bottom line that seems to make everything
pleasingly consistent (although not necessarily correct, of course!) is that
the most ``visible'' galaxies reside within the most massive dark matter halos,
and that generally speaking the star formation rate is proportional to the halo
mass. We believe that this kind of result provides strong empirical justification for the
general application of semi-analytic models which treat star formation as a function
of the parent dark matter halo properties using physically--motivated ``recipes'' (Baugh \et 1998,
Kauffman, Nusser, \& Steinmetz 1998). It is possible that further direct comparison of
the models to the observations could provide a means of fine-tuning the star formation prescriptions.

Regardless of the degree to which one is willing to believe that the observations and
theory are now pointing in the same direction, it is certain to be the case that
considerable progress in our understanding of the very-much-intertwined questions
of galaxy formation and the development of large scale structure will be made
in the immediate future. While at some level it is a bit of a disappointment that
galaxy clustering at high redshift is not telling us unambiguously about the background
cosmology, it certainly is the case that the observations can provide important tests
of our collective ideas about how, where, and when galaxies form relative to the  
dark matter distribution. It may well be that the relative simplicity in interpretation 
allowed by observations at very high redshift will more than make up for the difficulty
of obtaining the data. 

\section{Acknowledgments}

Much of the work described would not have been possible without the generous gift
from the W. M. Keck Foundation that allowed the construction of the Keck Observatory,
and the many people involved in building and supporting the telescopes
and the Low Resolution Imaging Spectrograph. 
This work has been financially supported by the US National Science foundation (CS, KA, MK) and
by grant HF-01071.01-94A from the Space Telescope Science Institute (MG).

\thebibliography{}
\item Adelberger, K. L., Steidel, C. C., Giavalisco, M., Dickinson, M., Pettini, M., \& Kellogg, M. 1998,
ApJ, in press.
\item
Bagla, J. S. 1998, MNRAS, in press. 
\item 
Baugh, C.M., Cole, S., Frenk, C.S., \& Lacey, C.G. 1998, ApJ, 498, 504
\item 
Bardeen, J. M., Bond, J. R., Kaiser, N., \& Szalay, A.S. 1986, ApJ, 304, 15
\item
Broadhurst, T., Ellis, R. S., Koo, D., \& Szalay, A. 1990, Nature, 343, 726
\item
Carlberg, R.G., Cowie, L. L., Songaila, A., \& Hu, E. M. 1997, ApJ, 484, 538
\item
Cohen, J.G., Hogg, D.W., Pahre, M.A., \& Blandford, R. D. 1996, ApJ, 462, L9
\item
Cole, S. \& Kaiser, N. 1989, MNRAS, 237, 1127
\item
Cowie, L. L., Songaila, A., Hu, E. M., \& Cohen, J. G. 1996, AJ, 1112, 839
\item
Dickinson, M. 1998, in The Hubble Deep Field, ed. M. Livio, M. Fall, \& P. Madau, (Cambridge: CUP), in press
\item 
Eke, V. R., Cole, S., \& Frenk, C. S. 1996, MNRAS, 282, 263
\item
Giavalisco, M. 1998, in The Hubble Deep Field, ed. M. Livio, M. Fall, \& P. Madau, (Cambridge: CUP), in press
\item
Giavalisco, M., Steidel, C. C., Adelberger, K. L., Dickinson, M. E., Pettini, M., \& Kellogg, M. 1998a, ApJ, in press
\item
Giavalisco, M. \et 1998b, in preparation
\item
Governato, F., Baugh, C. M., Frenk, C. S., Cole, S., Lacey, C. G., Quinn, T., \& Stadel, J. 1998,
  Nature, 392, 359 
\item
Jing, Y. P. \& Suto, Y.  1998, ApJL, submitted
\item
Kaiser, N.  1984, ApJL, 284, L9
\item
Kauffmann, G., Nusser, A., \& Steinmetz, M. 1997, MNRAS, 286, 795
\item
Le F\`evre, O., Hudon, D., Lilly, S. J., Crampton, D., Hammer, F., \& Tresse, L. 1996, ApJ, 461, 534
\item
Madau, P. 1995, ApJ, 441, 18
\item 
Madau, P., Ferguson, H.C., Dickinson, M., Giavalisco, M., Steidel, C.C., \& Fruchter, A. 1996, MNRAS, 283, 1388
\item
Mo, H. J., \& Fukugita 1996, ApJ, 467, L9
\item
Mo, H. J., \& White, S. D. M., 1996, MNRAS, 282, 347
\item
Oke, J. B. et al. 1995, PASP 107, 3750
\item
Peacock, J. A., \& Dodds, S. J. 1994, MNRAS, 267, 1020
\item
Pettini, M., Kellogg, M., Steidel, C. C., Dickinson, M., Adelberger, K. L., \& Giavalisco, M. 1998, ApJ, in press. 
\item
Press, W. H. \& Schechter, P. 1974, ApJ, 187, 425
\item
Steidel, C. C., Adelberger, K. L., Dickinson, M., Giavalisco, M.,
Pettini, M. \& Kellogg, M. 1998a, ApJ, 492, 428
\item
Steidel, C. C., Adelberger, K. L., Dickinson, M., Giavalisco, M., Pettini, M., \& Kellogg, M. 1998b, in
The Young Universe, eds. A. Fontana \& S. D'Odorico (San Francisco: ASP), in press.
\item
Steidel, C. C., Giavalisco, M., Pettini, M., Dickinson, M., \&
Adelberger, K. L. 1996, ApJ, 462, L17
\item
Steidel, C. C., Pettini, M., \& Hamilton, D. 1995, AJ, 110, 2519
\item
Steidel, C. C., \& Hamilton D. 1992, AJ, 104, 941 
\item
Steidel, C. C., \& Hamilton, D. 1993, AJ, 105, 2017
\item
Wechsler, R. H., Gross, M. A. K., Primack, J. R.,
Blumenthal, G. R. \& Dekel, A. 1998, ApJ, submitted
\item
White, S. D. M. \& Rees, M. J.  1978, MNRAS, 183, 341
\endthebibliography

\end{document}